\documentclass[final,5p,times,twocolumn]{elsarticle}

\usepackage{graphicx}
\usepackage{bm}
\usepackage{booktabs}
\usepackage{array}
\usepackage{hyperref}
\usepackage{cleveref}

\usepackage[pscoord]{eso-pic}

\newcommand{\placetextbox}[3]{
  \setbox0=\hbox{#3}
  \AddToShipoutPictureFG*{
    \put(\LenToUnit{#1\paperwidth},\LenToUnit{#2\paperheight}){\vtop{{\null}\makebox[0pt][c]{#3}}}%
  }%
}%

\usepackage{caption}
\usepackage{subcaption}

\usepackage{setspace}

\usepackage{lineno}

\journal{Journal for Radiation Physics and Chemistry}
\begin{document}

\begin{frontmatter}

\title{Exploring Large-Volume GAGG Scintillators for use in MeV Gamma-Ray Astrophysics}

\author[3]{Daniel Shy}
\author[3]{Richard Woolf}
\author[3]{Bernard Phlips}
\author[3]{Mary Johnson-Rambert}
\author[2]{Emily Kong}

\address[3]{U.S. Naval Research Laboratory, 4555 Overlook Ave SW, Washington, DC 20375, United States of America}
\address[2]{Technology Service Corporation, Arlington, VA,
22202, United States of America}


\begin{abstract}

Gamma-ray astrophysics in the MeV band is an exciting field in astronomy due to its potential for multi-messenger astrophysics. It has, however, remained under-explored when compared to other wavelengths. One reason for this observational gap is the difficulties with measuring these high-energy photons and the requirement of large amounts of detection material. In this work, we investigate the usage of large-volume GAGG scintillators for use as a calorimeter in future MeV telescopes. We developed a $5\times5$ array calorimeter utilizing $1\times1\times6 \ \mathrm{cm}^3$ GAGG crystals with onsemi C-series SiPM readout. We tested the calorimeter at the High Intensity Gamma-ray Facility (HIGS) with monoenergetic beams ranging from $2-25 \ \mathrm{MeV}$. Finally, we also investigate larger $1\times1\times8 \ \mathrm{cm}^3$ crystals and characterize their response across their depth when their surface treatment is either polished or frosted.

\end{abstract}


\end{frontmatter}


\section{Introduction}
\label{sec1}

\placetextbox{0.5}{0.05}{\large\textsf{DISTRIBUTION STATEMENT A. Approved for public release: distribution is unlimited.}}%


The gamma-ray astrophysical sky in the MeV range remains largely under-explored. The most recent large-scale mission dedicated to measuring and imaging the MeV gamma-ray sky was the Compton Gamma-Ray Observatory (CGRO), which was de-orbited in 2000~\cite{cgro}. Since then, the community has proposed several Compton and/or pair-production instruments to address what is known as the `MeV gap'. Most pair-production instruments are based on a two-subsystem design: a tracker to trace the movement of the cascade of particles from the gamma-ray interaction and a calorimeter to capture the ensuing electromagnetic showers. This tracker-calorimeter design is utilized by current missions such as the Fermi Space Telescope-LAT~\cite{FERMI} and AGILE~\cite{AGILE}. Missions that are in the concept and development phase include AMEGO-X~\cite{AMEGO-X}, ASTROGAM~\cite{ASTROGAM}, SMILE-2+~\cite{SMILE}, VLAST~\cite{VLAST}, and GECCO~\cite{GECCO}. 

This work investigates using gadolinium aluminum gallium garnet (GAGG) based scintillator to act as a high-energy calorimeter for potential usage in gamma-ray telescopes. GAGG presents several advantages over CsI, as outlined in Table~\ref{tab:GAGG}. First, it has superior energy resolution. Next, it is significantly denser than most other scintillators, which results in a smaller instrument without compromising the high-energy response. A faster decay time also allows for operation at higher count rates. Moreover, some reports declare a lower light-yield degradation due to radiation damage~\cite{Yoneyama_2018}. The table also reveals several disadvantages. Perhaps the most significant downside is its afterglow properties in that it is larger than most scintillators and endures for longer~\cite{gaggAfterglow,Yoneyama_2018}.

Although GAGG is a relatively new material, it does have a minor space heritage. Launched in December 2021, the GAgg Radiation Instrument (GARI) 1A and 1B~\cite{GARI} have been operating for several years~\cite{GARI_orbit}. Next, there have been several space-based instruments proposed that utilize GAGG: GALI~\cite{GALI}, HERMES~\cite{HERMES}, LISSAN~\cite{LISSAN}, GRAPE~\cite{GRAPE}, and FGS~\cite{FGS}. In addition, several studies have measured the internal activation and degradation associated with high-energy proton exposure~\cite{GAGG_Damage,DILILLO202233,koreanGAGG}.

\begin{table}[h!]

\centering
\caption{Some scintillator characteristics comparing GAGG and CsI:Tl. The bold text signifies the better crystal for the given property.}
\label{tab:GAGG}
\begin{tabular}{l|l|l}
\toprule

                            & GAGG~\cite{kinheng}               & CsI:Tl\cite{csi}      \\
                            \hline
                            \hline
Light Yield (ph/keV)        & \textbf{60}                 & 54 \\
Resolution (FWHM @ 662 keV) & \textbf{$\sim$5\%} & $\sim$8\%   \\
Density (g/$\mathrm{cm}^3$) & \textbf{6.6}       & 4.51        \\
Decay Time (ns)             & \textbf{$\leq$150} & 1000    \\
Afterglow~\cite{gaggAfterglow,Yoneyama_2018}             & Significant & \textbf{Minimal}    \\

\bottomrule
    
\end{tabular}
\end{table}

The outline of this manuscript is as follows: Sec.~\ref{sec:cal} describes the development and construction of a $5\times5$ array GAGG calorimeter utilizing $1 \times 1 \times 6 \ \mathrm{cm}^3$ crystals. Sec.~\ref{sec:HIGS} presents the calorimeter's response to high-energy gamma rays ranging from 2 MeV to 25 MeV. Finally, Sec.~\ref{sec:doi} investigates longer crystals $(1 \times 1 \times 8 \ \mathrm{cm}^3)$ with different surface treatments and their response as a function of depth.~\ref{sec:simAppendix} presents the simulated response of the $5 \times 5$ array calorimeter to high-energy gamma rays.

\section{Development of a $5\times5$ array GAGG Calorimeter}
\label{sec:cal}

\begin{figure}[h!]
  \centering
  \includegraphics[trim={0cm 0cm 0cm 0cm}, clip, width=\linewidth]{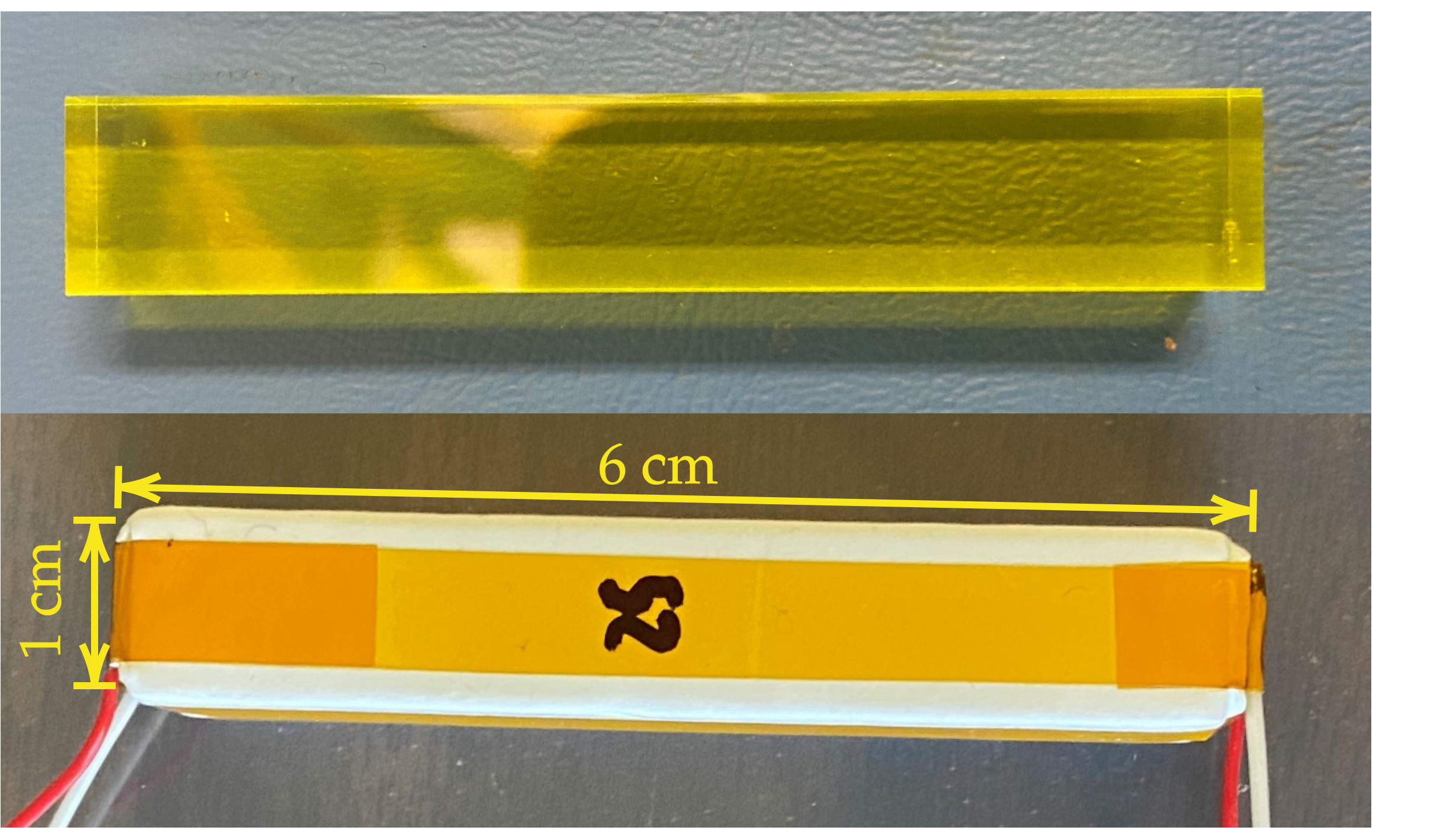}
  \caption{(Top) shows the bare unwrapped GAGG crystal while (Bottom) shows the fully wrapped and assembled crystal detector element. This crystal in this picture has a volume of $1 \times 1 \times 6 \ \mathrm{cm}^3$. }
  \label{fig:expSetup}
\end{figure}

In this work, we utilize high light-yield GAGG:Ce from Kinheng. The prototype in this section uses $1\times1\times6 \ \mathrm{cm}^3$ crystals. Each crystal is wrapped twice with a $216 \ \mu\mathrm{m}$ thick diffuse white reflector (Tetratex)~\cite{tetratex}. It was chosen based on our previous experience and its overall favorable reflective properties~\cite{csi, ieeeCsI}. We placed a $6\times6 \ \mathrm{mm}^2$ C-series SiPM by onsemi~\cite{cSiPM} on each end of the crystal. The C-series SiPM was also chosen due to our familiarity with it from the NeRDI-1A instrument~\cite{nerdi}. Fig.~\ref{fig:expSetup} shows the bare crystal (top) and a fully built crystal element (bottom).

\begin{figure}
     \centering
     \begin{subfigure}[b]{0.23\textwidth}
          \centering
         \includegraphics[trim={4cm 1.7cm 4cm 2.5cm}, clip,height=0.8\textwidth]{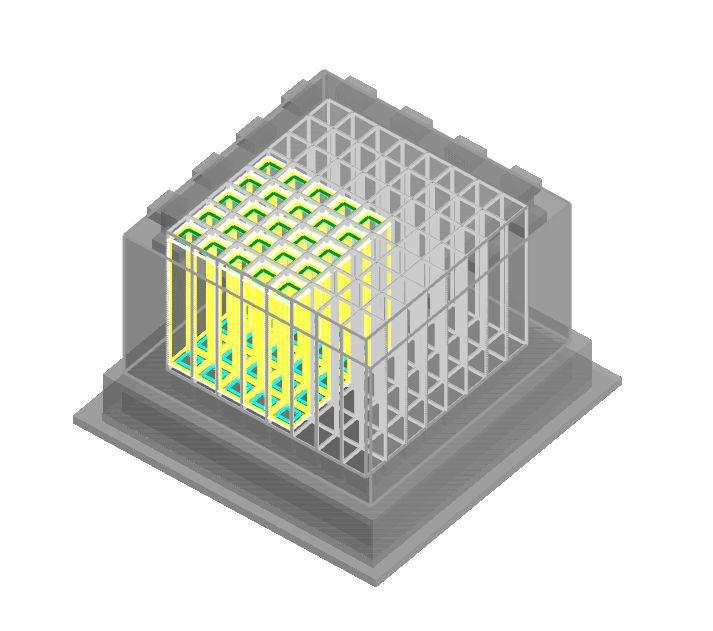}
         \caption{CAD Model}
         \label{fig:CAD}
     \end{subfigure}
      \begin{subfigure}[b]{0.23\textwidth}
         \centering
         \includegraphics[trim={6cm 1cm 6cm 1cm}, clip,height=0.8\textwidth]{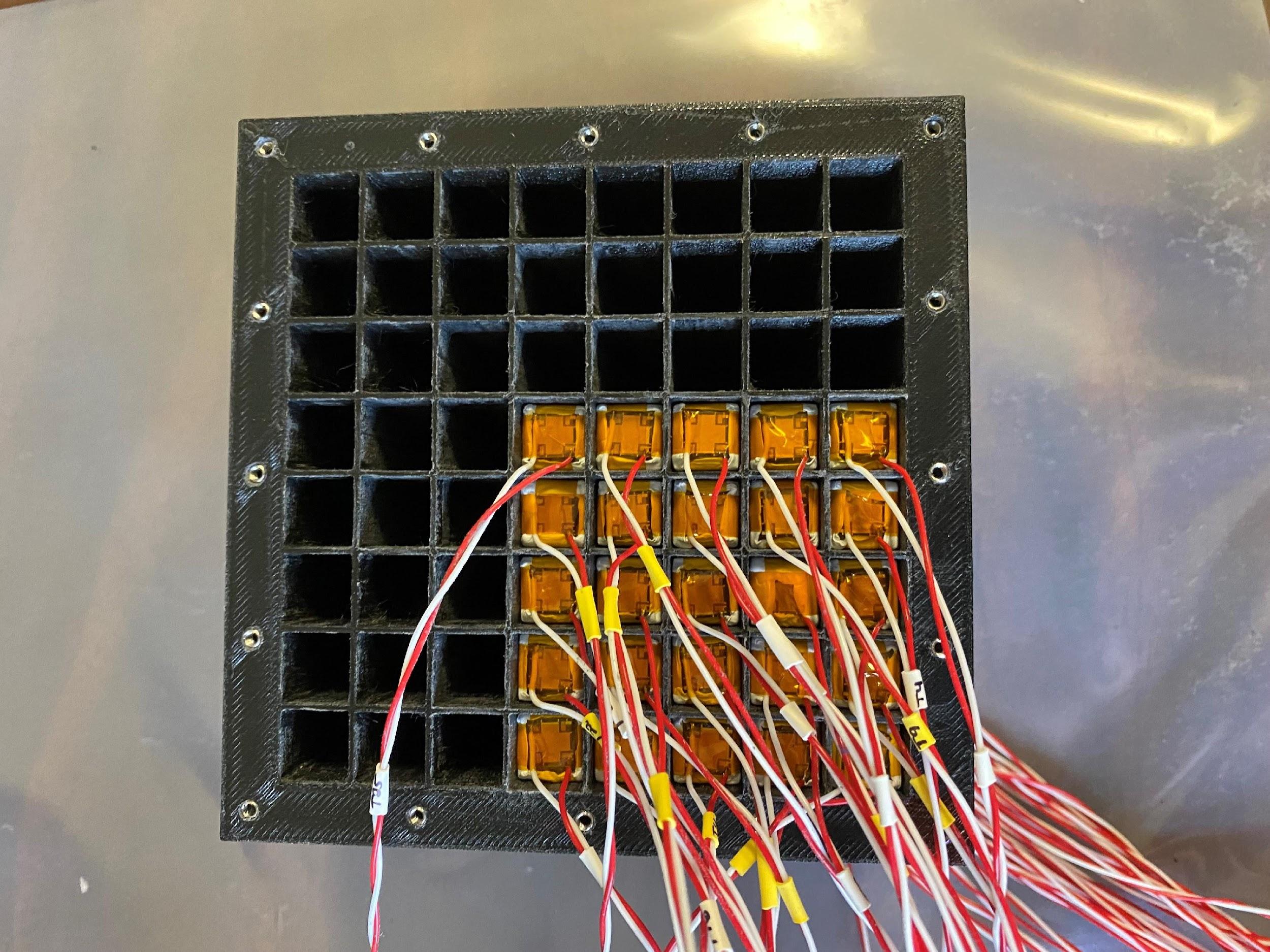}
         \caption{Top view of the calorimeter}
         \label{fig:TopPic}
     \end{subfigure}

        \caption{Images of the $5\times5$ array GAGG calorimeter}
        \label{fig:calorimeter}
\end{figure}

\begin{figure}[h!]
  \centering
  \includegraphics[trim={3cm 2cm 3cm 3cm}, clip, width=\linewidth]{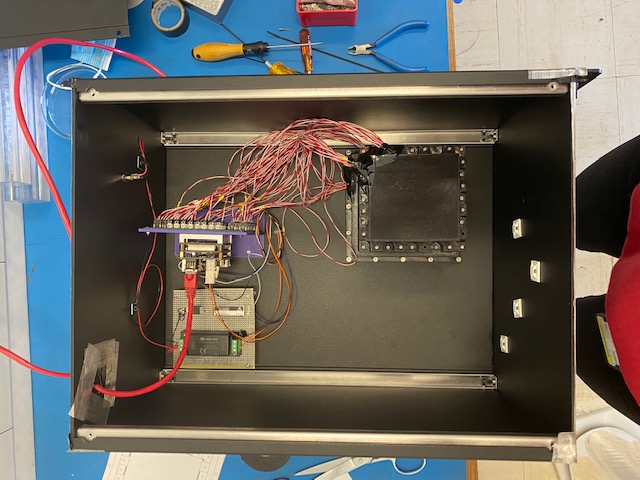}
  \caption{The $5\times5$ array calorimeter is placed inside the test box. The calorimeter array is housed in the upper right (black, 3D printed structure), while the ROSSPAD readout is placed on the left. A custom power supply, placed on the lower left, is utilized as a legacy from a different system that requires a higher current limit.}
  \label{fig:gaggBox}
\end{figure}

Since the crystals are not long enough to develop a hodoscopic design (as done with the Fermi-LAT~\cite{FERMI}), we instead opt for a `finger' based design and place the crystals upright and length-wise along the axis of the telescope. The crystals are housed in a 3D printed fixture arranged in a $5 \times 5$ array such that the center-to-center pitch is $1.1 \ \mathrm{cm}$. Fig.~\ref{fig:CAD} shows the computer-aided design (CAD) model of the calorimeter, while Fig.~\ref{fig:TopPic} shows the top view of the calorimeter. All SiPMs were read out by an IDE AS ROSSPAD~\cite{rosspad}. The entire system is then placed in a box to maintain light tightness, shown in Fig.~\ref{fig:gaggBox}. The general implementation, front-end, and analysis techniques are duplicated from a past prototype of the ComPair CsI subsystem~\cite{ComPairCsI}.

\subsection{Gamma-ray response}

The deposited energy is calculated as the geometric mean between each SiPM on each end for a given crystal:

\begin{equation}
\label{eq:energy}
E = \sqrt{Q_{top} \times Q_{bottom}},
\end{equation}

\noindent where $Q_{top}$ and $Q_{bottom}$ are the measured signals from the top and bottom SiPM on the same crystal. Although different energy treatments exist, we utilize this form based on our experience from the ComPair CsI balloon instrument~\cite{csi}.
 
\begin{figure}[h!]
  \centering
  \includegraphics[trim={0cm 0cm 0cm 0cm}, clip, width=\linewidth]{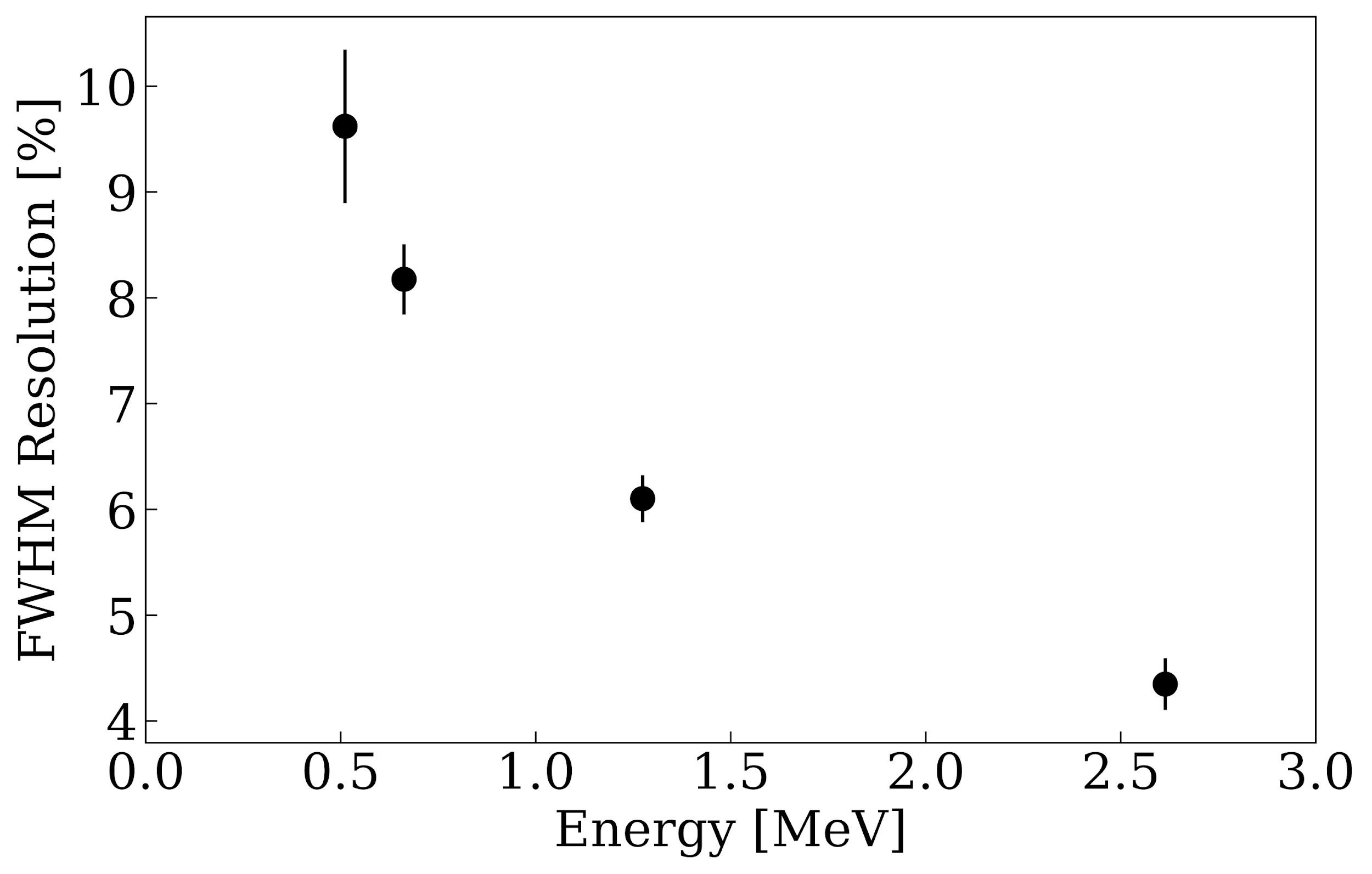}
  \caption{The average FWHM resolution for a given energy for all crystals with the error bars representing the observed standard deviation between the crystals. The sources we used included $^{22}\mathrm{Na}, ^{137}\mathrm{Cs}$ and Thoriated welding rods.}
  \label{fig:resolution}
\end{figure}

With the fully assembled calorimeter, we tested its response to $^{22}\mathrm{Na}, ^{137}\mathrm{Cs}$ and Thoriated welding rods. Fig.~\ref{fig:resolution} plots the percent full-width at half-maximum (FWHM) energy resolution as a function of the energy deposited. In the plot, only single crystal spectra are considered (i.e. coincidence between multiple crystals is ignored). The error bars represent the standard deviation between all the bars. 

The measured resolution at 662 keV is around 8\%, which is poorer than the general advertised value. This is most likely due to the difference in geometry as this GAGG is longer than mainstream crystals on the market. Moreover, GAGG is known to be significantly more self absorbing when compared to other scintillators~\cite{gaggAttenuation}. The degradation of scintillator resolution with geometry has been extensively documented~\cite{Shurcliff:49, 10.1063/1.1717121, BGOGeometry}.

\section{High-Energy Gamma-Ray Response of the Calorimeter}
\label{sec:HIGS}

\begin{figure}[h!]
  \centering
  \includegraphics[trim={0cm 0cm 10cm 0cm}, clip, width=\linewidth]{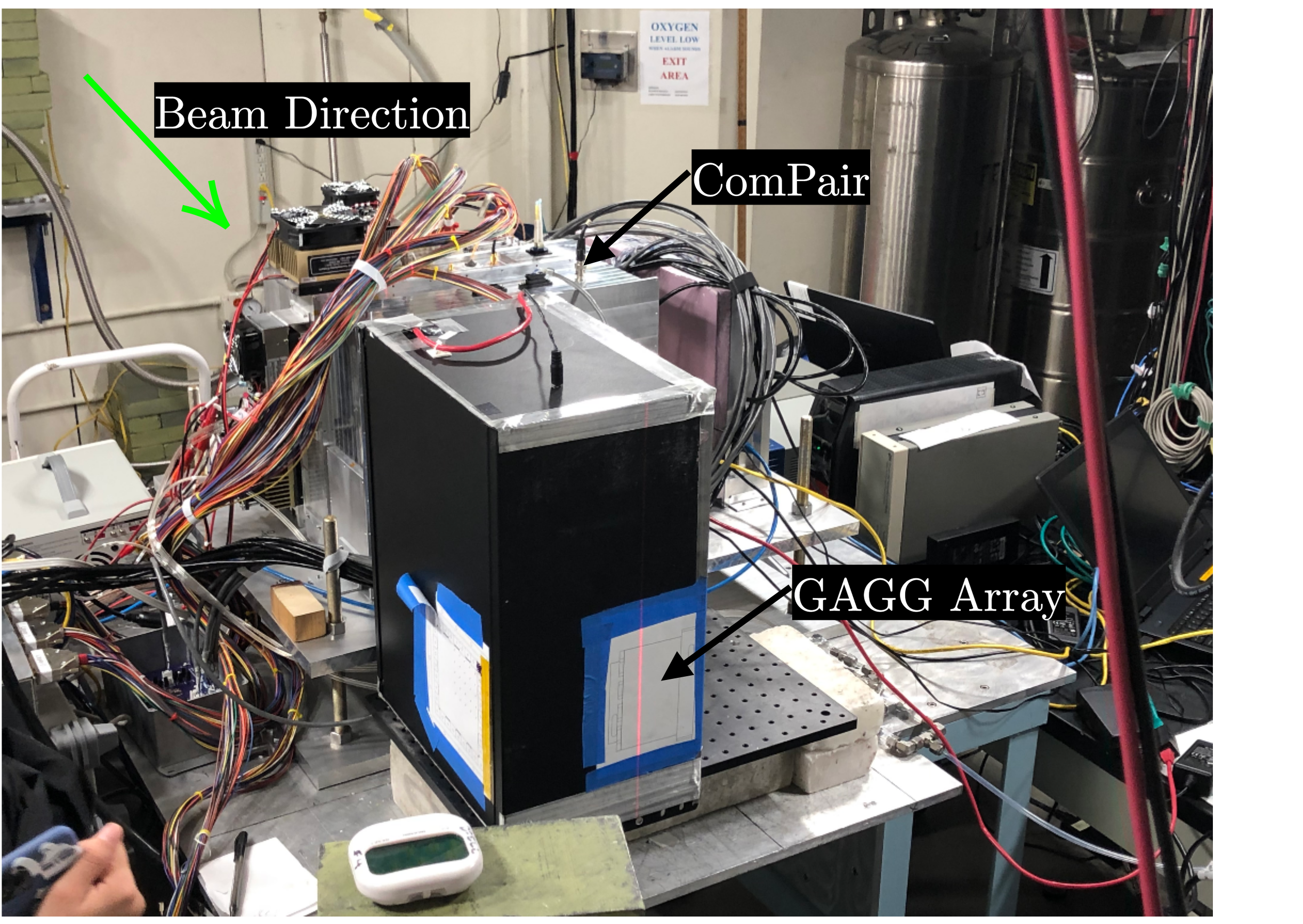}
  \caption{Experimental setup of the GAGG array conducted at the HIGS facility. The ComPair instrument is present and in front of the GAGG as this effort was conducted as a piggy back to their campaign.}
  \label{fig:expSetupHIGS}
\end{figure}

In April 2022, the calorimeter was taken to the Triangle Universities Nuclear Laboratory (TUNL) High Intensity Gamma-Ray Source (HIGS) facility. The facility can produce a monoenergetic gamma-ray beam with a diameter of $0.5 \ \mathrm{cm}$. The experiment utilized beam energies of 2, 5, 7, 15, and 25 MeV. Fig.~\ref{fig:expSetupHIGS} shows the experimental setup. Since this experiment was conducted as an add-on to the AMEGO/ComPair campaign~\cite{compair}, the GAGG calorimeter is placed downstream of the ComPair instrument. The ComPair instrument consists of 3 subsystems: 6 layers of $0.5 \ \mathrm{mm}$ of Si, $2 \ \mathrm{cm}$ of CZT, and $8.35 \ \mathrm{cm}$ of CsI.

\begin{figure}[h!]
  \centering
  \includegraphics[trim={0cm 0cm 0cm 0cm}, clip, width=\linewidth]{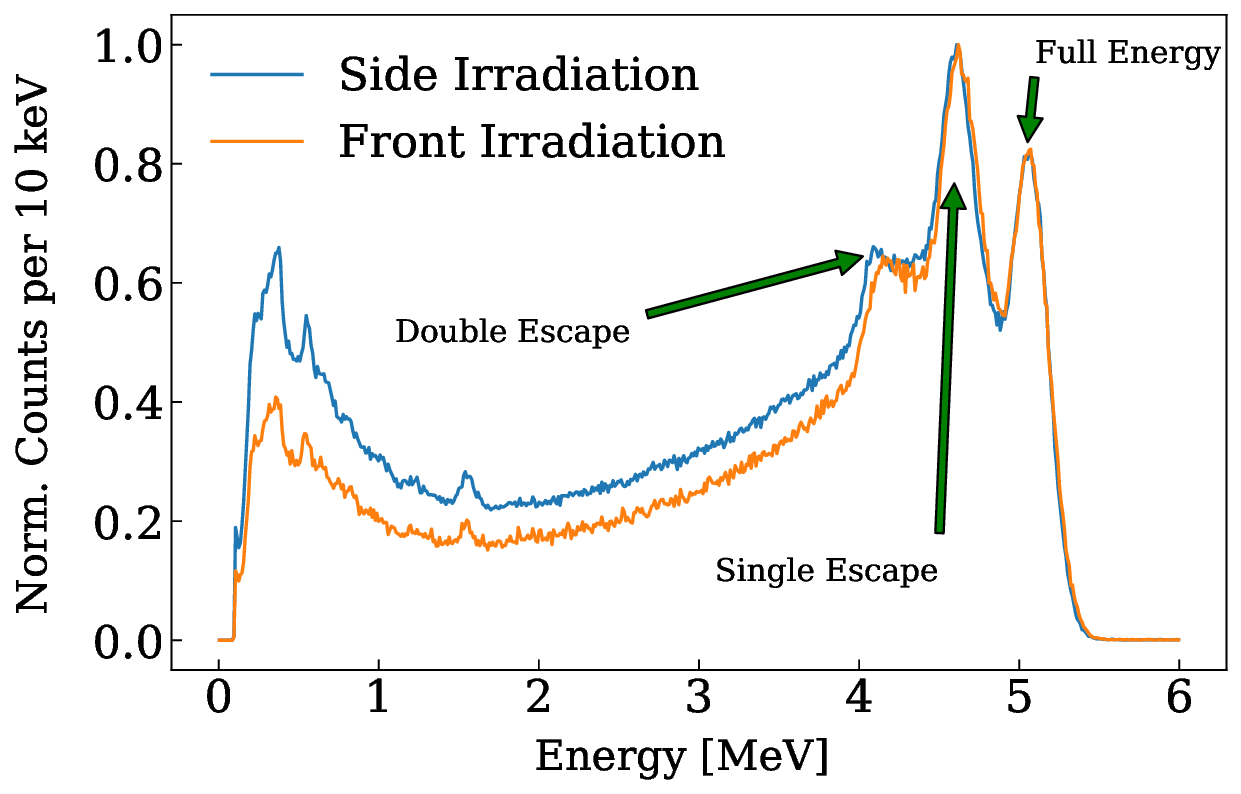}
  \caption{Full calorimeter response to a 5 MeV beam while the calorimeter is oriented at different poses.}
  \label{fig:5MeVSum}
\end{figure}

During the experiment, we placed the calorimeter in different positions and angles relative to the beam. This is to change the interaction profile of the gammas with the calorimeter. Fig.~\ref{fig:5MeVSum} shows the summed response to different irradiation configurations of a 5 MeV beam. Side irradiation refers to when the beam is shot into the center column of the calorimeter while front irradiation is where the beam is aimed into the $1\times1 \ \mathrm{cm^2}$ face of the center crystal in the array. We note that the difference in the full calorimeter response is relatively invariant in that the ratio between the escape peaks, continuum, and full-energy peak are very similar. This is likely due to the near-cubic geometry of the calorimeter which results in a uniform sensitivity to source location.

\begin{figure*}
     \centering
     \begin{subfigure}[b]{0.45\textwidth}
          \centering
         \includegraphics[trim={0cm 0cm 0cm 0cm}, clip,height=0.8\textwidth]{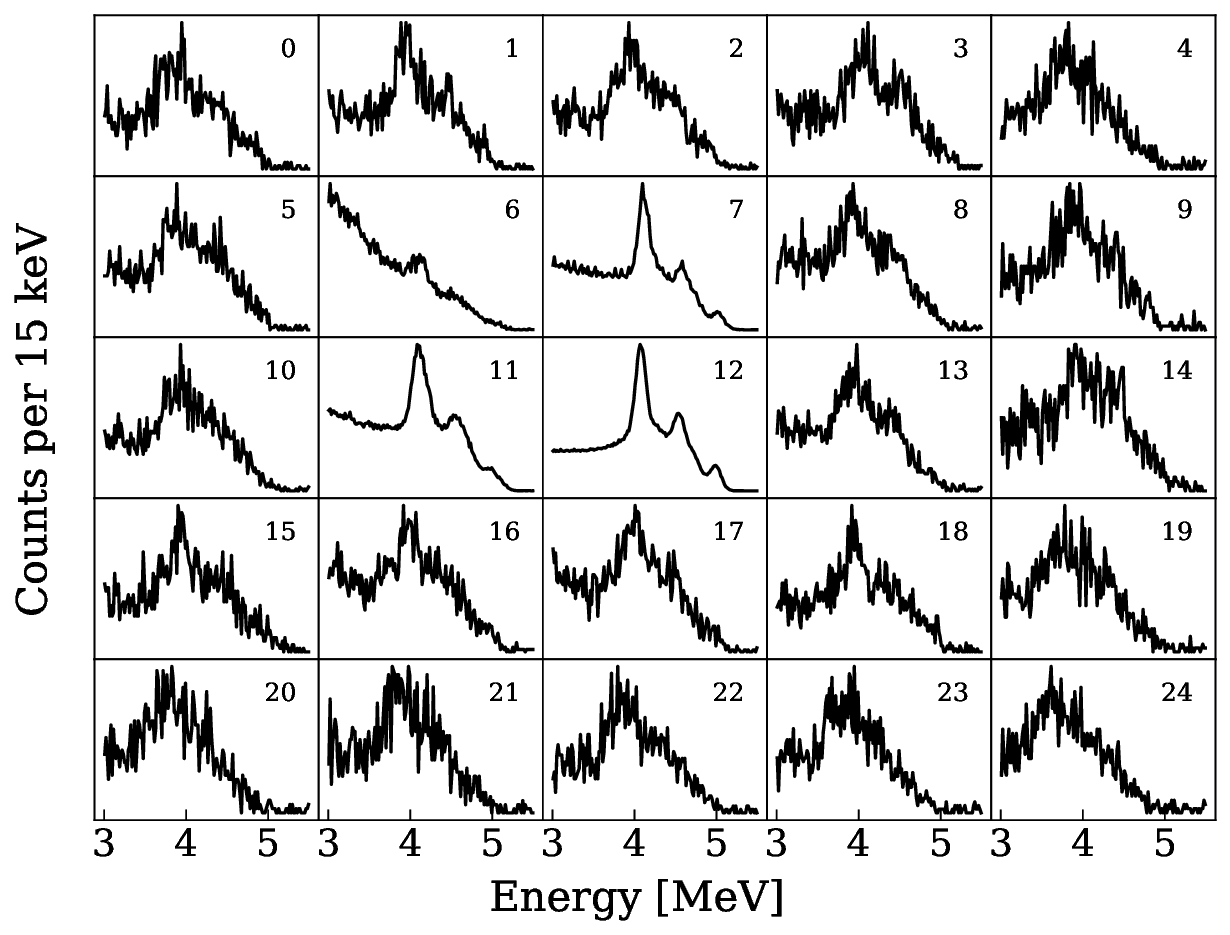}
         \caption{Face irradiation}
         \label{fig:intoPageProfile}
     \end{subfigure}
     \hfill
      \begin{subfigure}[b]{0.45\textwidth}
         \centering
         \includegraphics[trim={0cm 0cm 0cm 0cm}, clip,height=0.8\textwidth]{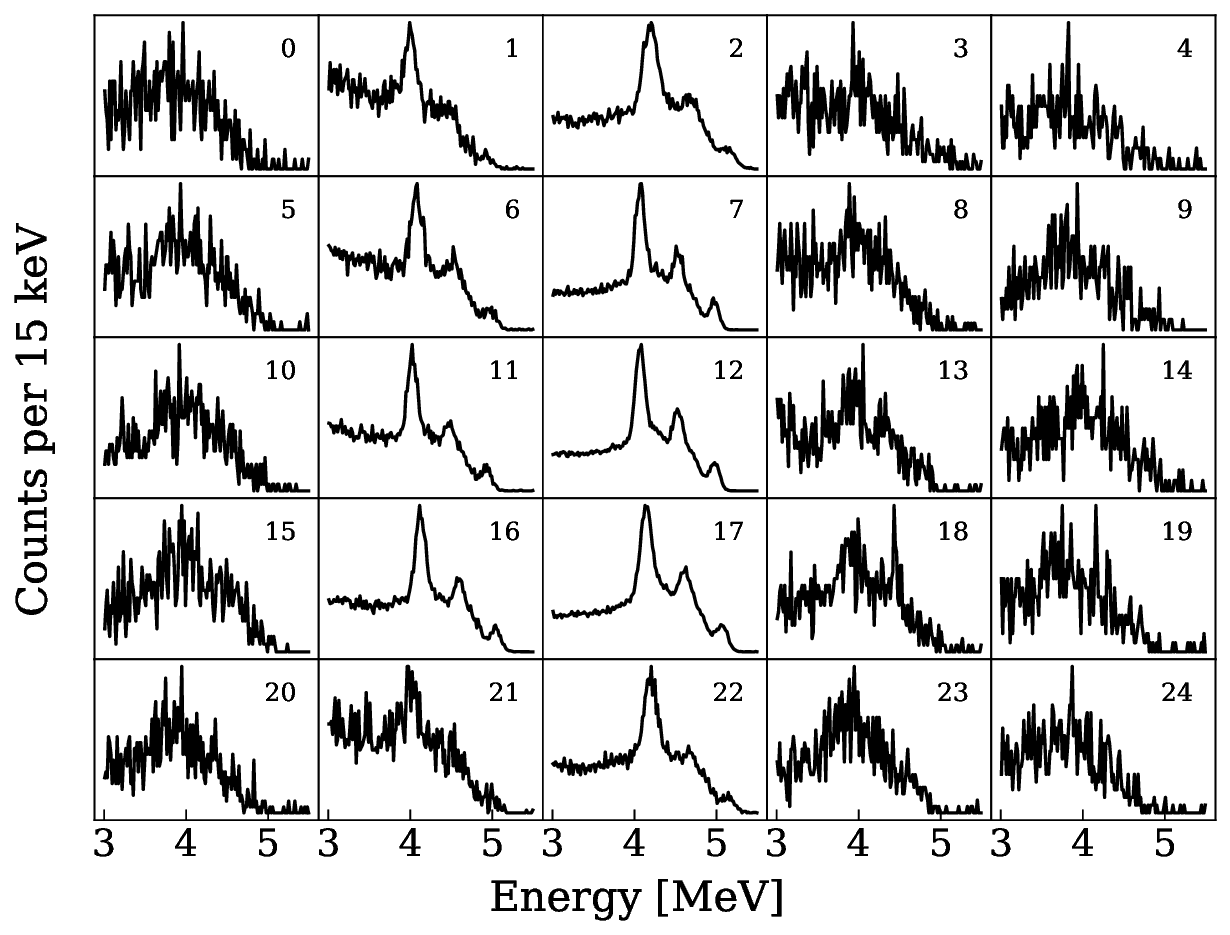}
         \caption{Side irradiation}
         \label{fig:sideIrradiationProfile}
     \end{subfigure}
        \caption{Crystal-to-crystal response to a 5 MeV beam while the calorimeter is oriented at different poses. The y-axes of the subplots are not shared with the range left floating. The reader can infer the number of counts in each subplot from the statistical noise in each trace.}
        \label{fig:interactionProfile}
\end{figure*}

Fig.~\ref{fig:interactionProfile} shows the crystal-by-crystal response to different irradiation configurations relative to the 5 MeV beam. Fig.~\ref{fig:intoPageProfile} plots the front irradiation while Fig.~\ref{fig:sideIrradiationProfile} plots side irradiation. The two irradiations show the expected full energy and escape peaks in the target crystals. The double escape peak is enhanced when viewing the spectrum from a single crystal compared to the summed response in Fig.~\ref{fig:5MeVSum}. More specifically, the double escape peak is the most prominent feature in the single crystal spectra. The summed response has the single escape peak as the most prominent with the full-energy peak being a close 2nd. The surrounding crystals contain significant continuum and 511 keV depositions from the ensuing cascades. The high energy response in the surrounding crystals most likely originates from the off-axis artifacts of the beam.

\begin{figure}[h!]
  \centering
  \includegraphics[trim={0cm 0cm 0cm 0cm}, clip, width=\linewidth]{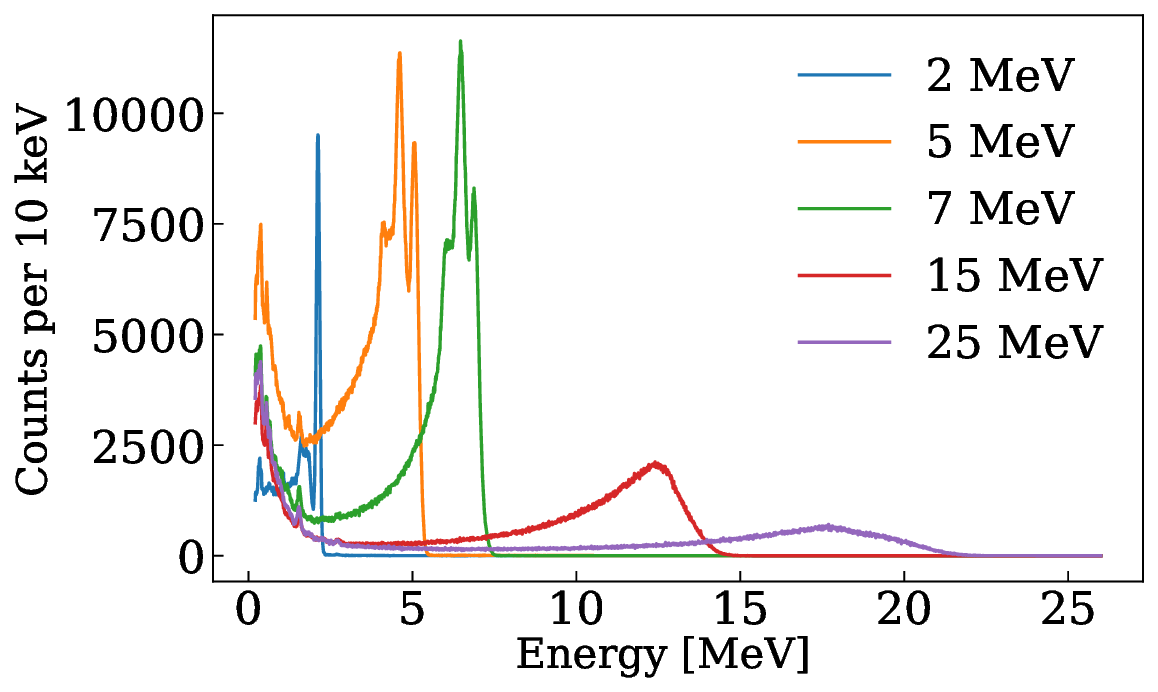}
  \caption{GAGG Calorimeter response to monoenergetic gamma rays ranging from $2 \ \mathrm{MeV} -25 \ \mathrm{MeV}$}
  \label{fig:higsSpectra}
\end{figure}

Fig.~\ref{fig:higsSpectra} plots all the acquired spectra from the HIGS campaign. Full energy peaks are clearly visible for the 2, 5, and 7 $\mathrm{MeV}$ beams. Also featured are the single and double escape peaks. However, runs with 15 and 25 $\mathrm{MeV}$ do not present any evidence of escape peaks, likely due to the low full-energy efficiency and low full containment of the ensuing spallation. The experimental response matches the simulated one, presented in~\ref{sec:simAppendix}.

\section{Depth of Interaction response study of GAGG Crystals}
\label{sec:doi}

\begin{figure}[]
  \centering
  \includegraphics[trim={2cm 4cm 1cm 2cm}, clip, width=\linewidth]{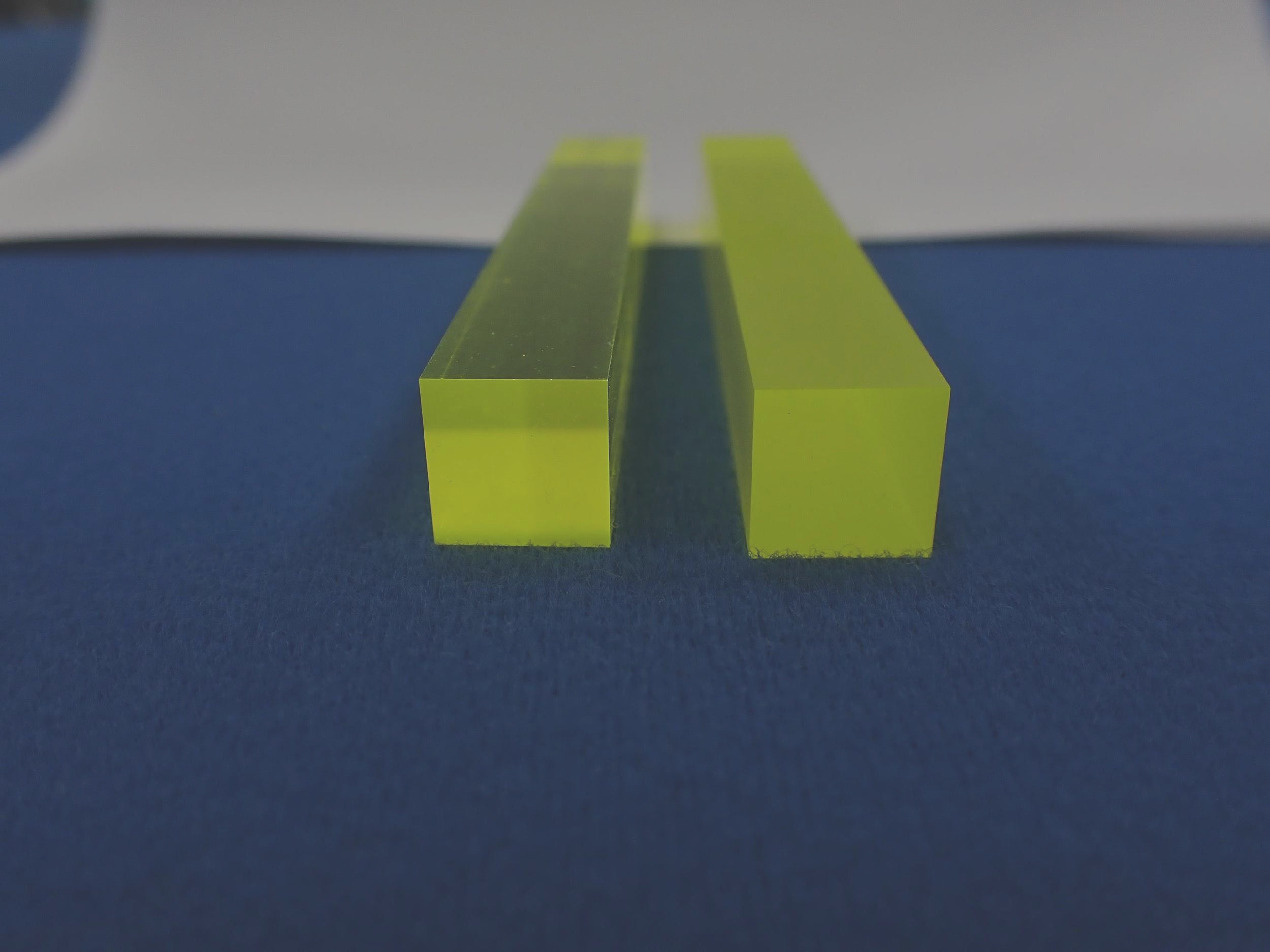}
  \caption{Picture of the $1 \times 1 \times 8 \ \mathrm{cm}^3$ crystals that are either polished (left) or frosted on all sides (right).}
  \label{fig:bare8Crystals}
\end{figure}

We expand our investigation of GAGG by exploring longer crystals that are $1 \times 1 \times 8 \ \mathrm{cm}^3$ in volume. Moving to the longer crystals is of interest as it increases the detection efficiency and full-energy containment. In Fermi's LAT calorimeter, two of the four sides were roughened to introduce a depth-dependent light output when read out on each end of the crystal~\cite{GLAST_Update}. In this section, we investigate the difference in depth of interaction response when the sides of the crystals are either polished or frosted. Fig.~\ref{fig:bare8Crystals} shows the roughened crystal (on the right), next to a polished version (left). Each crystal is packaged and read out like the crystals described in Sec.~\ref{sec:cal}.

\subsection{Collimated Depth Measurements}

\begin{figure}[h!]
  \centering
  \includegraphics[trim={0cm 0cm 0cm 0cm}, clip, width=\linewidth]{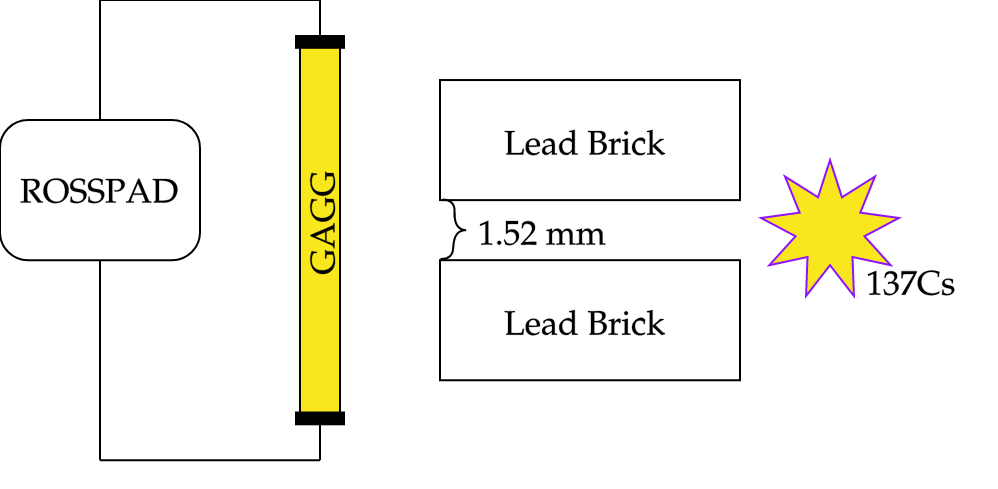}
  \caption{Diagram of the experimental setup to measure the depth response of the $1 \times 1 \times 8 \ \mathrm{cm}^3$ GAGG crystal.}
  \label{fig:collimatorDiagram}
\end{figure}

In this section, we conducted collimated measurements at each depth to verify the scintillation properties across the entire crystal and the depth reconstruction scheme. A fan beam is produced by using two lead bricks separated by a $1.52 \ \mathrm{mm}$ thick American penny with a $^{137}\mathrm{Cs}$ source placed on the other end. The assembly is mounted on a robotic gantry that is programmed to move in $5 \ \mathrm{mm}$ steps and scan along the length of the GAGG crystal. Fig.~\ref{fig:collimatorDiagram} illustrates the experimental setup.

\begin{figure}[h!]
  \centering
  \includegraphics[trim={0cm 0cm 0cm 0cm}, clip, width=\linewidth]{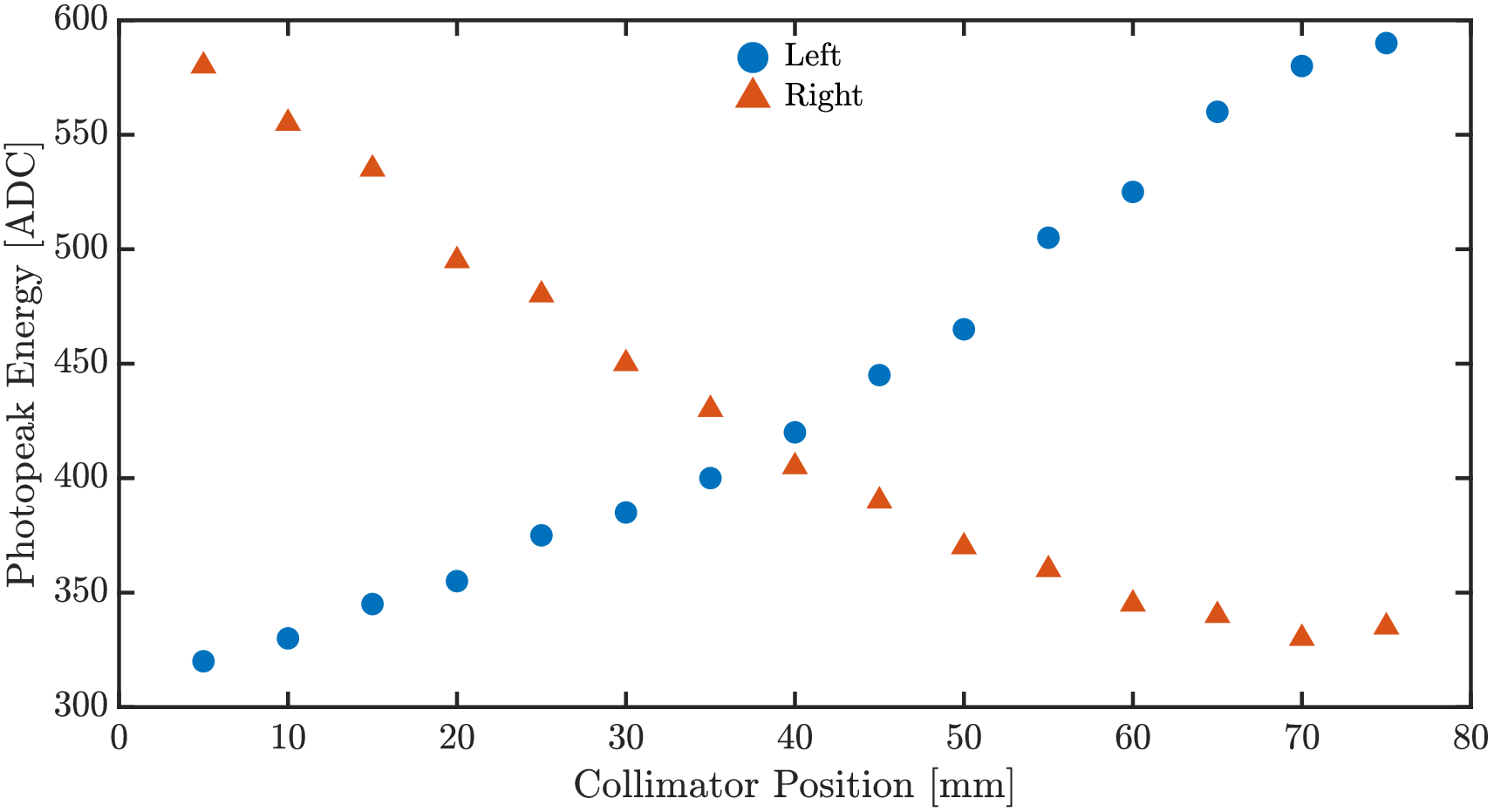}
  \caption{Photopeak response of each SiPM to collimated $^{137}\mathrm{Cs}$ source placed at different depth along the crystal. }
  \label{fig:collimatorStudySiPM}
\end{figure}

Fig.~\ref{fig:collimatorStudySiPM} plots the $^{137}\mathrm{Cs}$ photopeak response of the SiPM on each end. We observe that at the center of the crystal ($40 \ \mathrm{mm}$), the signals are not perfectly equal. This could arise from asymmetric gain between the two ends or the difference between the optical coupling of the SiPM to the crystal. Most of the crystal displays a linear response along the depth. Finer steps are required to be taken to observe the edge effect when the interactions are close to one end of the crystal.

We calculate the depth of interaction (DOI) using Eq.~\ref{eq:doi}.

\begin{equation}
\label{eq:doi}
\mathrm{DOI} = \frac{Q_{top}-Q_{bottom}}{Q_{top}+Q_{bottom}}.
\end{equation}

\begin{figure}[h!]
  \centering
  \includegraphics[trim={0cm 0cm 0cm 0cm}, clip, width=\linewidth]{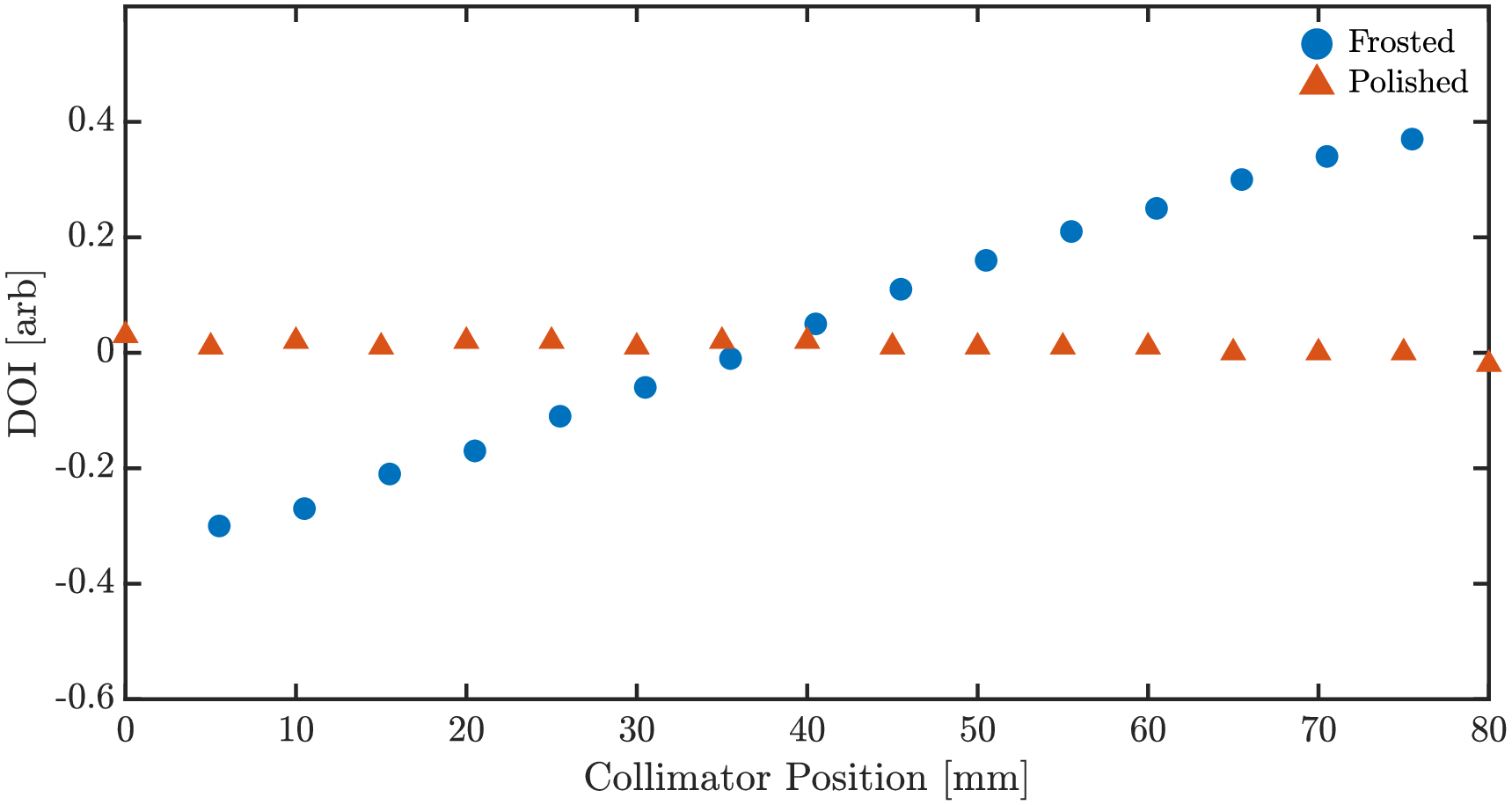}
  \caption{ Calculated DOI as a function of collimator location for a crystal that was either polished or frosted on each end.}
  \label{fig:collimatorStudyDOI}
\end{figure}

We can then make a DOI calculation as a function of the collimator position and plot it in Fig.~\ref{fig:collimatorStudyDOI}. The figure also plots the calculated DOI for a polished crystal. The frosted crystal displays a very linear response to the different depths indicating that the applied surface roughening is sufficient. On the other hand, the polished crystal displays a negligible amount of depth response. Nevertheless, both crystals display a fully active volume in that their entire length was sensitive to gamma rays.

\subsection{Flood Irradiation Response}

\begin{figure}[h!]
  \centering
  \includegraphics[trim={0cm 0cm 0cm 0cm}, clip, width=\linewidth]{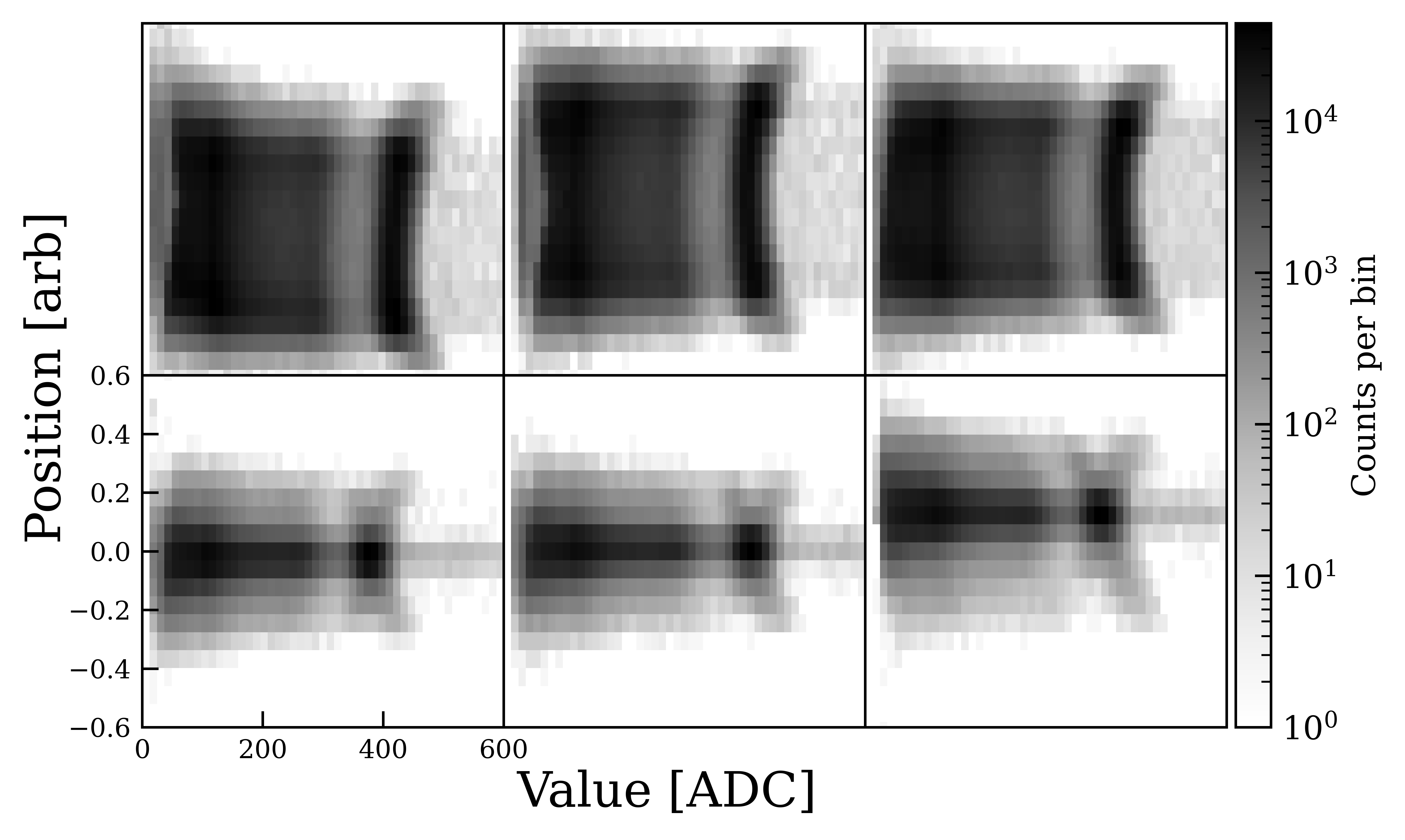}
  \caption{Raw depth-energy response to a $^{137}\mathrm{Cs}$ source. The top row plots the 3 different frosted crystals while the bottom row plots the 3 responses of the polished crystals.}
  \label{fig:rawDepthEnergy}
\end{figure}

This section demonstrates the response of 3 frosted and 3 polished crystals to a $^{137} \mathrm{Cs}$ flood irradiation. Using Eq.~\ref{eq:energy} and~\ref{eq:doi}, we can develop a bivariate histogram to study the depth-energy response. Fig.~\ref{fig:rawDepthEnergy} plots the depth-energy response of the frosted (top row) and polished crystals (bottom row). Once again, the polished crystals display no depth response while the frosted ones show a depth response. In the frosted crystals, the photopeak is visible along the multiple depths. The DOI value range is not centered around 0 for all crystals, likely due to the asymmetric gain between each end.

\begin{figure}[h!]
  \centering
  \includegraphics[trim={0cm 0cm 0cm 0cm}, clip, width=\linewidth]{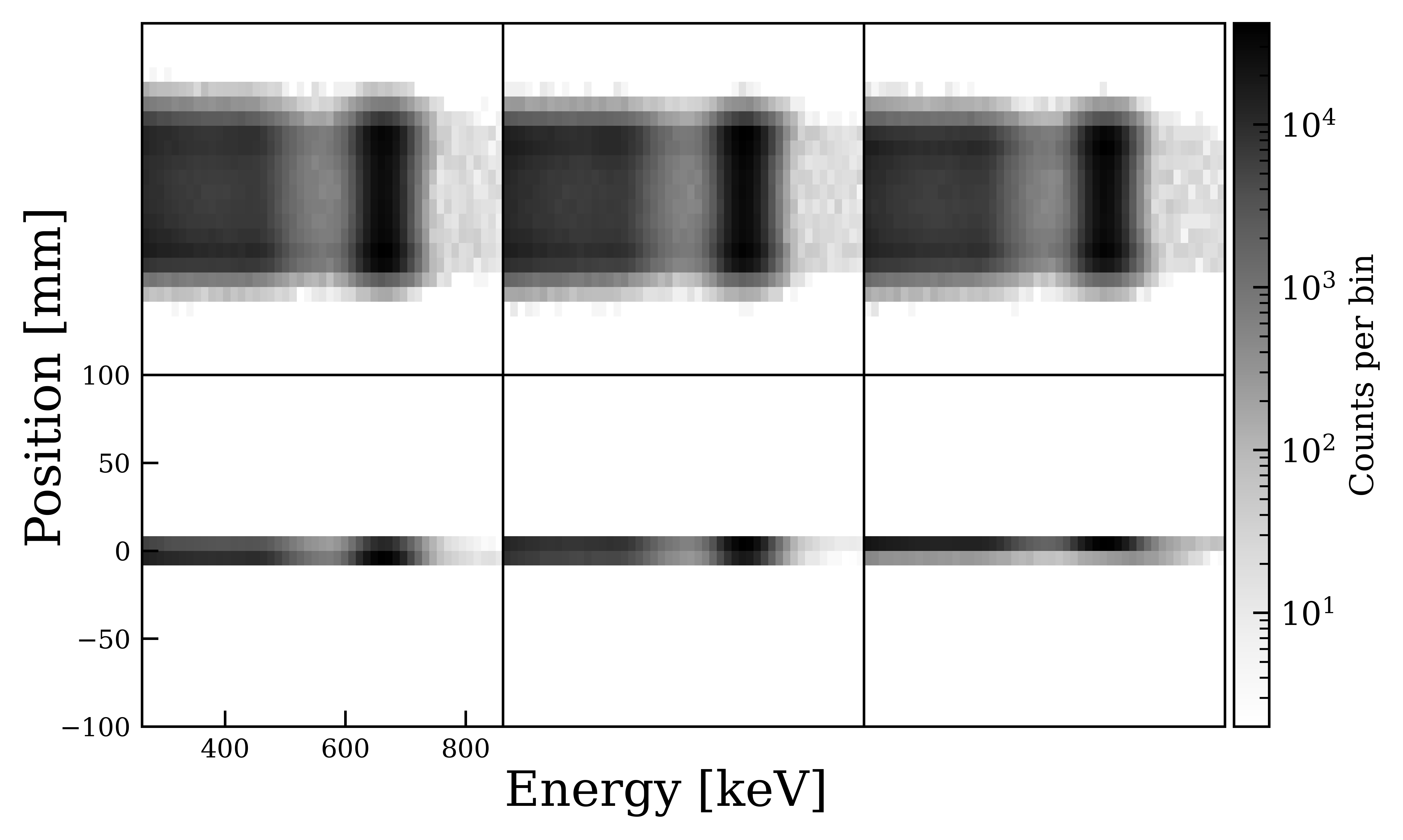}
  \caption{Corrected depth-energy response of the $1 \times 1 \times 8 \ \mathrm{cm}^2$ GAGG crystals. The top row plots the 3 different frosted crystals while the bottom row plots 3 polished crystals.}
  \label{fig:correctedDepthEnergy}
\end{figure}

The energy and depth are then calibrated using techniques described in~\cite{ComPairCsI}. Fig.~\ref{fig:correctedDepthEnergy} plots the corrected energy-depth response, highlighting the $^{137}\mathrm{Cs}$ peak, where the top and bottom rows plot the frosted and polished crystals respectively.

\begin{figure}[h!]
  \centering
  \includegraphics[trim={0cm 0cm 0cm 0cm}, clip, width=\linewidth]{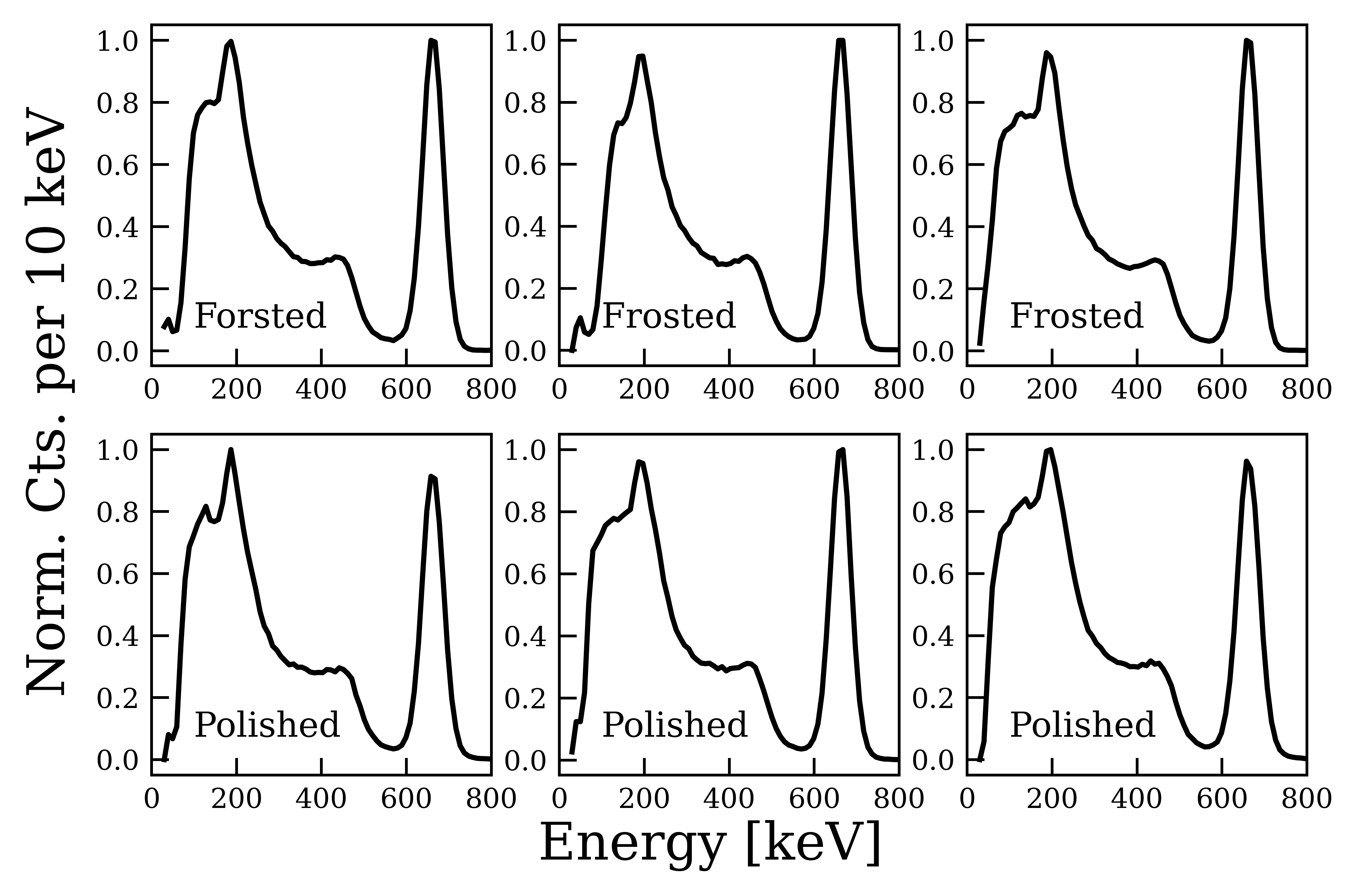}
  \caption{Energy response of the $1 \times 1 \times 8 \ \mathrm{cm}^2$ GAGG crystals using the calibration matrix. The top row plots the 3 different frosted crystals while the bottom row plots 3 polished crystals.}
  \label{fig:correctedEnergy}
\end{figure}

We plot the spectra of the crystals in Fig.~\ref{fig:correctedEnergy} to highlight their gamma-ray response. We show that the depth response of the frosted crystals is removed. Due to the crystal geometry, the full-energy peak-to-continuum ratio is rather poor. Table~\ref{tab:8mmResolution} summarizes the resolution of each crystal. We show that the resolution is slightly degraded when compared to the $6 \ \mathrm{cm}$ crystal. A possible explanation of why the frosted crystals outperform the polished ones is that the depth corrections can correct any regional defects in the crystal. However, a larger sample of crystals is required to declare a conclusive statement.

\begin{table}[h!]

\centering
\caption{Percent full width at half maximum at $662 \ \mathrm{keV}$ displaying the resolution of the $8 \ \mathrm{cm}$ long crystals.}
\label{tab:8mmResolution}
\begin{tabular}{l|c|c}
\toprule

                & Best Crystal [\%]      & Average [\%]     \\
                \hline
                \hline
Frosted			& 8.2			& 8.5	$\pm$ 0.2	\\
Polished		& 8.5			& 8.8	$\pm$ 0.3  \\

\bottomrule
    
\end{tabular}
\end{table}

\section{Conclusion and Future Work}

This work explores the usage of GAGG as a high-energy calorimeter for a MeV range gamma-ray telescope. We built a $5 \times 5$ array of GAGG crystals, each $1 \times 1 \times 6 \ \mathrm{cm}^3$ in size. This system was then characterized with monoenergetic gamma rays ranging from $2$ to $25 \ \mathrm{MeV}$. The results show good agreement with simulations and present an initial proof of concept. 

Next, we investigate the depth response of $1 \times 1 \times 8 \ \mathrm{cm}^3$ crystals that are either frosted or polished on all sides. Collimated measurements show that the longer $8 \ \mathrm{cm}$ crystals are fully active showing that the long GAGG crystals are sensitive to gamma rays throughout the entire volume. Moreover, polishing the crystals on all 4 sides resulted in the desirable depth-dependent signals when the fully polished crystal yielded negligible depth response. The polished crystals displayed a near-linear response depth response when using Eq.~\ref{eq:doi} to calculate the depth, reducing the need for advanced calibration and non-linearity corrections. Future iterations of this finger calorimeter design would opt for crystals where the 4 long faces are polished. The depth response is desired when attempting to reconstruct the particle tracks or ensuring electromagnetic showers for the targeted gamma-ray telescope applications.

We are currently developing a larger $8 \times 8$ array calorimeter using the $1 \times 1 \times 8 \ \mathrm{cm}^3$ crystals. In addition, we are developing SiPM readouts that field two different SiPM species such that their different gains will the dynamic energy range of the readout~\cite{dualGainSiPM}. The `dual-gain' SiPMs are currently under fabrication by Fondazione Bruno Kessler~\cite{amegoXCalIEEE}.

\section*{Acknowledgment}

This work is supported by the Office of Naval Research 6.1. We are grateful for the efforts of the staff at the HIGS facility to ensure a successful experiment.

We thank the AMEGO/ComPair team for the opportunity to join their HIGS Campaign. ComPair is funded under NASA Astrophysics Research and Analysis (APRA) grants NNH14ZDA001N-APRA, NNH15ZDA001N-APRA, NNH18ZDA001N-APRA, NNH21ZDA001N-APRA.

\appendix

\section{Simulated Response of the GAGG Calorimeter}
\label{sec:simAppendix}

We modeled the $5\times5$ GAGG calorimeter using the SWORD simulation package~\cite{SWORD}. In addition to the calorimeter (shown in Fig.~\ref{fig:CAD}), the ComPair system was also modeled to simulate the HIGS beam test conditions. The plots in this section show the GAGG calorimeter's response to $2, 5, 7, 15$ and $25 \ \mathrm{MeV}$ monoenergetic gamma rays. The simulated energy blurring is accomplished using the mean energy resolution presented in Fig.~\ref{fig:resolution}. However, we did not account for the crystal-to-crystal variations in energy response.

\begin{figure}[h!]
  \centering
  \includegraphics[trim={1cm 15cm 6cm 0cm}, clip, width=\linewidth]{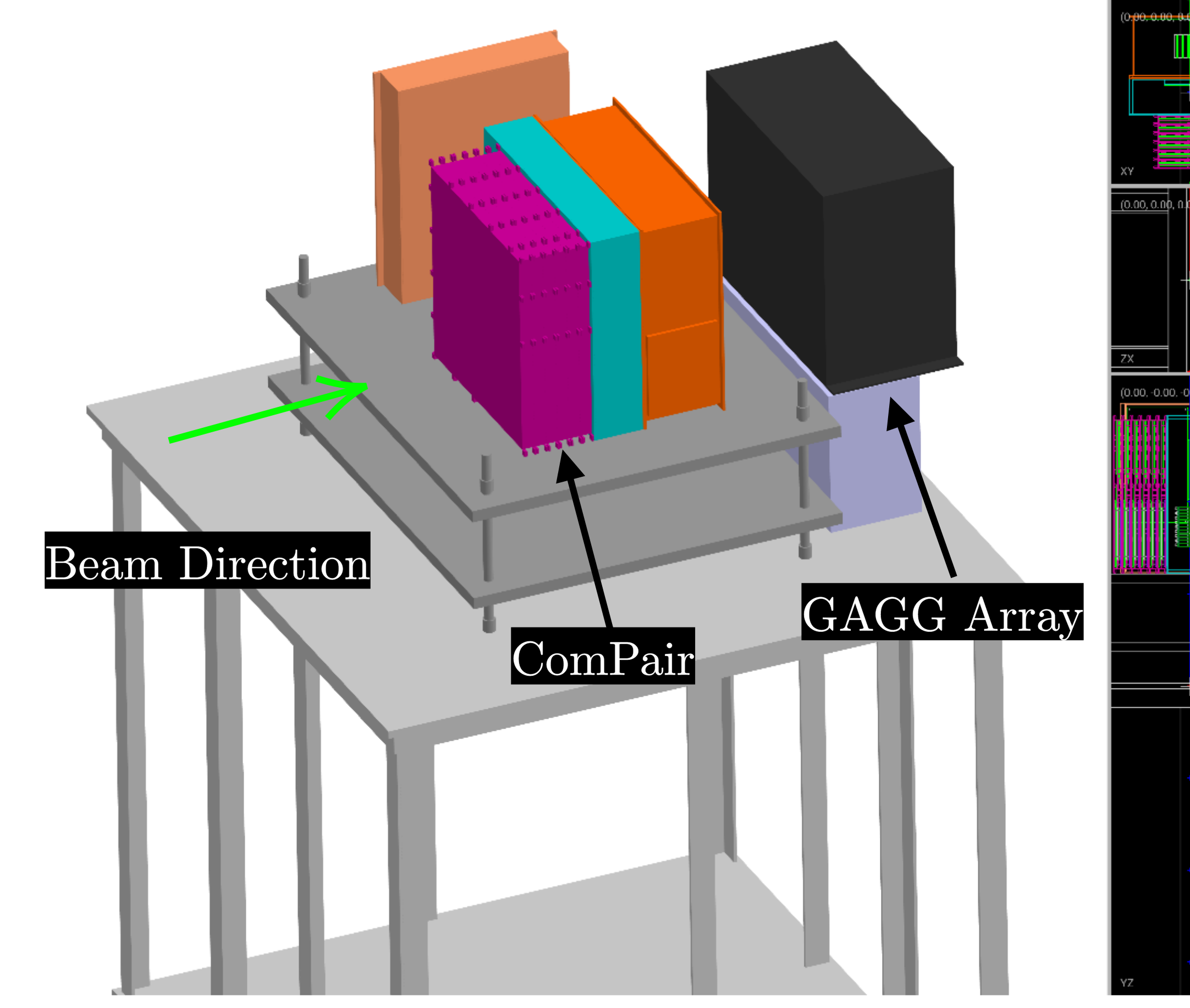}
  \caption{CAD model of the simulated environment and instrument mass model for the experimental setup taken place at HIGS.}
  \label{fig:HIGS_SWORD}
\end{figure}

For energies ranging from $2-7 \ \mathrm{MeV}$, a clear full energy peak is visible along with the two escape peaks, which mirrors what is observed during experimentation. The $15 \ \mathrm{MeV}$ simulated run shows very modest full energy and escape peaks. On the other hand, the experimental result shows a skewed left continuum with no obvious features. This is likely a result of the simulation not accounting for the crystal-to-crystal variations. The $25 \ \mathrm{MeV}$ run shows a continuum with no spectral peaks, similar to the experimental run.

\begin{figure}[h!]
  \centering
  \includegraphics[trim={0cm 0cm 0cm 0cm}, clip, width=\linewidth]{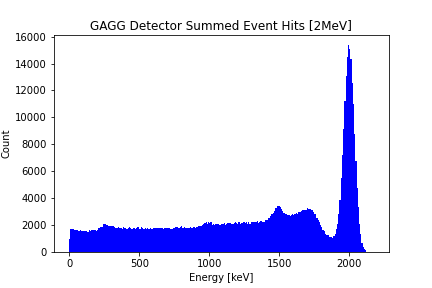}
  \caption{Simulated response of the 5x5 array GAGG calorimeter to 2 MeV gamma rays.}
  \label{fig:2MeVSim}
\end{figure}

\begin{figure}[h!]
  \centering
  \includegraphics[trim={0cm 0cm 0cm 0cm}, clip, width=\linewidth]{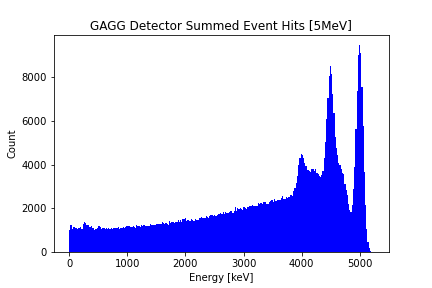}
  \caption{Simulated response of the 5x5 array GAGG calorimeter to 5 MeV gamma rays.}
  \label{fig:5MeVSim}
\end{figure}

\begin{figure}[h!]
  \centering
  \includegraphics[trim={0cm 0cm 0cm 0cm}, clip, width=\linewidth]{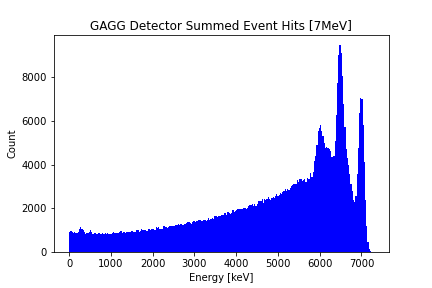}
  \caption{Simulated response of the 5x5 array GAGG calorimeter to 7 MeV gamma rays.}
  \label{fig:7MeVSim}
\end{figure}

\begin{figure}[h!]
  \centering
  \includegraphics[trim={0cm 0cm 0cm 0cm}, clip, width=\linewidth]{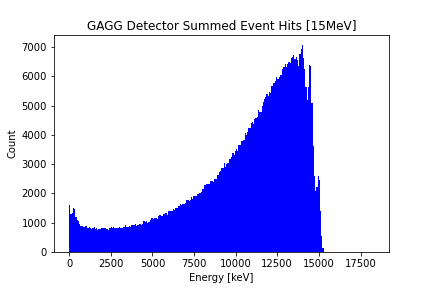}
  \caption{Simulated response of the 5x5 array GAGG calorimeter to 15 MeV gamma rays.}
  \label{fig:15MeVSim}
\end{figure}

\begin{figure}[h!]
  \centering
  \includegraphics[trim={0cm 0cm 0cm 0cm}, clip, width=\linewidth]{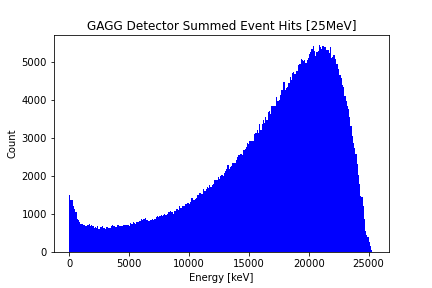}
  \caption{Simulated response of the 5x5 array GAGG calorimeter to 25 MeV gamma rays.}
  \label{fig:25MeVSim}
\end{figure}

\clearpage

\nocite{*}
\bibliographystyle{elsarticle-num}

\bibliography{IEEEbib}

\end{document}